\newcommand{\epst}{\tilde{\epsilon}_1}
\newcommand{\lp}{\lambda_{ab}}
\newcommand{\xhat}{{\bf \hat{x}}}
\newcommand{\yhat}{{\bf \hat{y}}}
\newcommand{\zhat}{{\bf \hat{z}}}

\documentstyle[prl,aps,epsf,twocolumn]{revtex}
\begin{document}
\draft
\tighten

\title{Low-Temperature Vortex Dynamics in Twinned Superconductors}

\author{M.Cristina Marchetti}
\address{Physics Department, Syracuse University, Syracuse, NY 13244}

\author{Valerii M. Vinokur}
\address{Argonne National Laboratory, Materials Science Division, Argonne, 
IL 60439}

\date{\today}

\maketitle

\widetext
%\narrowtext
\begin{abstract}
We discuss the low-temperature dynamics of magnetic flux lines in 
samples with a family of 
parallel twin planes. A current applied along the
twin planes drives flux motion in the direction transverse to the
planes and acts like an electric field applied to {\it one-dimensional}
carriers in disordered semiconductors. 
As in flux arrays with columnar pins, there is a regime where the dynamics 
is dominated by superkink excitations
that correspond to Mott variable
range hopping (VRH) of carriers. In one dimension, however,
rare events, such as large regions void of twin planes, can impede
VRH and dominate transport in samples that are sufficiently long 
in the direction of flux motion. In short samples
rare regions can be responsible for mesoscopic effects.
\end{abstract}

\narrowtext
%\widetext

The static and dynamic properties of magnetic flux lines 
in copper-oxide superconductors are strongly affected by
pinning by point, linear and planar disorder \cite{tora}.
{\it Linear} correlated disorder in the form of
columnar defects produced by 
bombardment of the crystal with energetic heavy ions
has been shown to greatly enhance pinning
in both yttrium and thallium-based compounds \cite{civale}.
Twin boundaries are an example of {\it planar} disorder that is ubiquitous in
superconducting $YBa_2Cu_3O_{7-x}$ and $La_2CuO_4$.
Early decoration experiments indicated that
the superconducting order parameter
is suppressed at a twin boundary and the twin attracts 
the vortices \cite{dolan}.
Extensive investigations of twin-boundary pinning have been carried out
by Kwok and coworkers \cite{kwokone,kwoktwo}. 
These authors studied a variety of YBCO single
crystal samples containing single families of parallel twins lying
in planes spanned by the $c$ axis, with
spacings ranging from microns down to several hundred Angstroms.
Transport experiments \cite{kwokone} show
clear evidence of strong twin-boundary pinning
even in the flux liquid phase
for external fields along the $c$ axis and driving currents
in the $ab$ plane and parallel to the plane of the twins (resulting in
a Lorentz force normal to the twin planes). 
For this geometry
the linear resistivity 
drops sharply at a characteristic
temperature where twin-boundary pinning sets in
\cite{kwokone}.
In addition, there is a sharp downward dip in the resistivity as a function of 
angle as the external field is rotated through the 
$\hat{c}$ direction \cite{kwoktwo}.

The static and dynamic properties of flux-line assemblies in the presence of 
a random array of {\it columnar pins} have been studied by mapping
the physics of magnetic flux lines onto the problem of localization
of quantum mechanical bosons in {\it two dimensions}
\cite{nvlong}. 
At low temperatures there is a ``Bose glass'' phase, with flux lines localized on columnar pins, separated by a phase transition from an entangled flux
liquid of delocalized lines.
\newline\vskip 1.445truein\noindent 
Transport in the Bose glass phase
closely resembles the variable-range hopping (VRH) of electrons in 
disordered semiconductors in two dimensions \cite{shklovskii}.

In this paper we study flux-line dynamics 
in the presence 
of a {\it single family of parallel twin boundaries} lying in planes containing
the $c$ axis, for $\vec{H}\parallel\hat{c}$.
We find
that due to the quasi-one-dimensional nature of vortex transport 
a new regime can arise
at low currents where flux-line dynamics is dominated by
rare events, such as large regions voids of twin planes, that can
be responsible for new mesoscopic phenomena.

At low temperatures, when the average vortex spacing $a_0\approx(\phi_0/B)$
exceeds the average distance $d$ between twin planes, 
all flux lines are localized 
by the pinning potential in the direction normal to the twins,
progressively ``filling'' the planar pins as the field is
increased. We only consider fields below 
$B_f\approx \phi_0/d^2$, where 
the flux lines fill the twins,
and neglect any additional weak point disorder in the sample.
We focus
on flux motion transverse to the twin planes,
which resembles the hopping of
electrons in {\it one-dimensional} disordered superconductors.
The current density in the usual hopping conductivity problem corresponds to
the vortex velocity (i.e., voltage) and the electrical conductivity 
maps onto the resistivity from vortex motion.
Transport in this low temperature and field regime is dominated by 
single-vortex
dynamics. 
Flux motion is described in terms of thermally activated jumps of
the vortices
over the relevant pinning energy barriers ${\cal U}(L,J)$, yielding
a resistivity $\rho={\cal E}/J$,
\cite{tora}
\begin{equation}
\label{eq:taff}
\rho(T)\approx \rho_0e^{-{\cal U}/T},
\end{equation}
where $\rho_0$ is a characteristic flux-flow resistivity.
In the following we determine the barrier heights ${\cal U}(L,J)$
(Table 1) corresponding
to various transport regimes and the boundaries between the
different regimes in the $(L,J)$ plane.

Our starting point is the model-free energy discussed in \cite{nvlong}
for flux lines defined by their
trajectories $\{{\bf r}_i(z)\}$ as they traverse a sample
of thickness $L$ with a
family of parallel twin boundaries
parallel to the $zx$ plane,  
\begin{equation}
\label{freemany}
F_{(N)}=\sum_{i=1}^N F_i + {1\over 2}\sum_{i\not= j}
 \int_0^L V(|{\bf r}_i(z)-{\bf r}_j(z)|) dz,
\end{equation}
where $V(r)$ is the intervortex interaction \cite{interaction}
and $F_i$ is the 
single-flux line free energy,
\begin{equation}
\label{freeone}
F_i=\int_0^L dz \bigg[{\epst\over 2}
  \Big({d{\bf r}_i(z)\over dz}\Big)^2
  +V_D(y_i(z))\bigg].
\end{equation}
Here $\epst$ is the tilt modulus ($\epst=(M_{\perp}/M_z)\epsilon_1<\epsilon_1$,
with $\epsilon_1$ the line tension) and
$V_D(y)$
represents a $z$-and $x$-independent pinning potential \cite{gesh}.
At low temperatures
we model $V_D(y)$ as an array of identical one-dimensional
square potential wells of depth $U_0$, width $2b_0$ and average spacing 
$d>>b_0$,
passing completely through the sample in the $x$ and $z$ directions and
centered at uniformly distributed 
random positions along the $y$ axis.

As discussed in \cite{nvlong} and \cite{drnrev}, 
many relevant results
regarding the statistical mechanics of flux lines can be obtained from 
elementary quantum mechanics by mapping the vortex trajectories
onto the imaginary time path integrals of
{\it two-dimensional} particles in a static random
potential $V_D(y)$.  In this mapping $k_BT$ plays the role of $\hbar$,
$\epst$ that of the mass of the fictitious quantum particle, $L^{-1}$ that
of the particle's temperature. 
If we assume that each flux line
spends most of its time near the attractive twin planes,
the dynamics of flux lines
driven by a Lorentz force normal to the twin planes
can be described by a tight-binding model for {\it one-dimensional}
bosons, where the lattice sites are
the positions of the twin planes.
Flux lines can ``tunnel'' 
between twins $i$ and $j$
separated by a distance $d_{ij}$ at a rate given by 
the tunneling matrix element
$t_{ij}\sim 2 U(T)e^{-E_{ij}/T}$,
where $E_{ij}=\sqrt{2\epst U(T)}d_{ij}$ is the energy of
a ``kink'' configuration connecting the two pins and
$U(T)$ is the effective binding free energy
per unit length of a flux line trapped near a
twin plane. 
At low temperature $U(T)\approx U_0$.
Thermal fluctuations renormalize $U_0$ above a crossover temperature,
as discussed in  \cite{drnrev}.
The modeling of the dynamics of {\it two-dimensional} 
bosons in terms of
a tight binding model in {\it one dimension} can be understood as
a result of vortex-vortex interactions which confine each flux line within
a ``cage'' of radius $\sim a_0$ formed by its neighbors.
Each line can be viewed
\begin{table}
\caption{Energy barriers determining the various contributions to
the resistivity of Eq. 1, with $\alpha=g(\mu)dU(T)$.}
\begin{tabular}{lrl}
Linear&Nonlinear\\
\tableline
${\cal U}_{rf}=UL=E_k(L/w_k)$ & ${\cal U}_{hl}=E_k(J_1/J)$\\
${\cal U}_{nnh}=E_k$& \\
${\cal U}_{Mott}=E_k(L/\alpha w_k)^{1/2}$& ${\cal U}_{VRH}=E_k(J_1/\alpha 
J)^{1/2}$\\
\end{tabular}
\label{table1} 
\end{table}

\noindent as moving within a one-dimensional
``channel'' of width $\sim a_0$.
Another effect of interactions is the energy cost
for an additional flux line to be placed on an already filled
twin boundary.
This is incorporated into an energy cost $V_0$ for double
occupancy of a site of the one-dimensional tight-binding lattice,
which is estimated as 
$V_0\approx 4\epsilon_0 d^2/a_0^2$ for
$T<T^*d/b_0[\ln(a_0/2d)]^{-1}$ and
$d<<a_0<<\lp$.

We consider vortex transport in the
presence of a driving current ${\bf J}\perp{\bf H}$ parallel to the twin 
planes, i.e., ${\bf J}=-J\xhat$. The applied current exerts a Lorentz
force per unit length on the vortices,
${\bf f}_L={\phi_0\over c}\zhat\times{\bf J}=
           \yhat f_L$
in the direction transverse to the twin planes.
In the context of the analogy with boson quantum mechanics,
this term represents a fictitious ``electric field'' 
${\bf E}={1\over c}\zhat\times{\bf J}=\yhat J/c$ acting on particles with 
``charge'' $\phi_0$.
At low temperatures the critical current
can be obtained by equating the Lorentz force 
to $U_0/b_0$, where $b_0\sim \xi_{ab}$, i.e., $J_c(0)\approx cU_0/\phi_0b_0$.
Renormalization of $J_c$ by thermal fluctuations 
have been discussed in \cite{nvlong}.

\begin{figure}
\epsfxsize=3.0truein
\hskip 0.5truein \epsffile{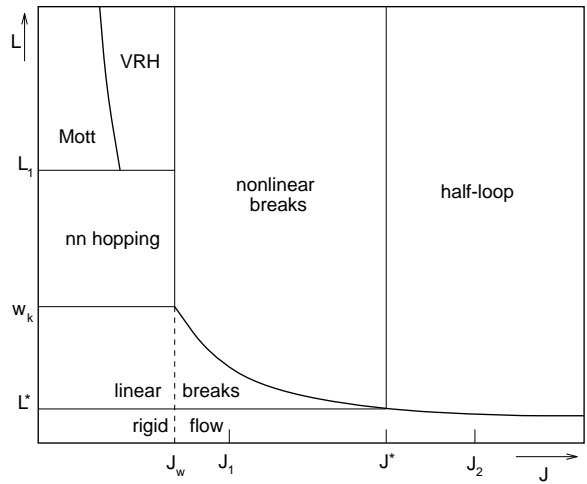}
\caption{The $(L,J)$ phase diagram for $\alpha=g(\mu)dU<1$.}
\label{phaseone}
\end{figure}

If we neglect rare events, the energy barriers determining the resistivity
of Eq. \ref{eq:taff} can be estimated as the saddle-point free energy
associated with the low-lying
excitations from the ground state (where all the lines are localized
on twins). The results are summarized in Table 1.
The phase diagram in the $(L,J)$ plane is controlled
by the parameter $\alpha=g(\mu)dU(T)$, where $g(\mu)$ is the density of states
for the most weakly bound flux lines and
$\mu\approx\phi_0( H-H_{c1})/4\pi$ is the chemical potential which 
fixes the flux line density.
The low-lying excitations that govern
transport at low temperature have been discussed
before in the context of flux arrays pinned by columnar defects \cite{nvlong}.
The only difference here is in the 
exponent of the barrier associated with
VRH, both in the linear and nonlinear regime. In the presence of linear
defects ${\cal U}_{VRH}\sim(\tilde{J}/J)^{1/3}$, while for planar
defect the corresponding exponent is $1/2$, as shown in Table 1.
The boundaries of the $(L,J)$ phase diagrams (Figs. 1 and 2)
have not been discussed before
in detail.
The line $J=J_L(L)$ defines the boundary in the $(L,J)$ plane that separates 
the regions of linear ($J<J_L(L)$) and nonlinear ($J>J_L(L)$) response,
with $J_L=cE_k/(\phi_0Ld)$ and $E_k=\sqrt{\epst U(T)}d$
the energy of a kink connecting neighboring pins separated by
$d$.
In the thermodynamic limit $J_L\rightarrow 0$ and the IV characteristic
is nonlinear at all currents.
In samples of very small thickness $L$ there is
a linear resistivity at low currents due to the flow of
flux-line segments of length $L$ and typical transverse width
$y_{rf}\approx dL/w_k$, with $w_k=E_k/U$
the width of a kink
connecting pins separated by $d$.
When $w_k<L<L_1$, where
$L_1=E_k/\gamma$ is the length below which dispersion from
tunneling and interactions can be neglected,
transport occurs via the
hopping of vortices between nearest neighbor (nn)
pinning sites. 
Here $\gamma$ is the width of the impurity band
arising from level dispersion from tunneling
and intervortex interaction.
The region of the phase diagram dominated by nn hopping is only
present if $L_1> w_k$, or $\gamma <U$
\cite{notenn}.
When $L>L_1$ the 
dispersion of energies between different pinning sites makes motion 
by nearest neighbor hopping energetically 
unfavorable. Tunneling
occurs instead via the formation of ``superkinks'' 
that throw a vortex
segment onto a spatially remote pin connecting 
states which optimize the tunneling probability. 
Tunneling via superkinks is the analogue of Mott variable-range hopping of electrons between
localized states in semiconductors. 
The corresponding energy barriers ${\cal U}_{Mott}$ is given in Table 1.

For $J>J_L(L)$ the resistivity is nonlinear. 
For $J_1<J<J_c$, with $J_1=cU(T)/\phi_0d$,
flux motion occurs via thermally 
activated ``half-loop'' configurations of transverse width
$y_{hl}\approx U/f_L$.
For $J<J_1$ the size of the transverse displacement of the liberated vortex 
segment exceeds the average distance $d$ between twin planes and transport
occurs via VRH which generalizes the Mott mechanism 
to the nonlinear case. 
The characteristic current scales that governs VRH is 
$J_0=J_1/\alpha$. 
The VRH contribution to the resistivity dominates
that from half loop only if
if ${\cal U}_{VRH}<{\cal U}_{hl}$, or $J<J_2=J_1\alpha$. 
All current scales in Fig. 1
are much smaller than the pair breaking current 
$J_{pb}=4c\epsilon_0/(3\sqrt{3}\phi_0\xi_{ab})$. 
The Mott and the rigid flow 
regimes are separated by a horizontal line above which 
${\cal U}_{Mott}<{\cal U}_{rf}$. Similarly, the condition 
${\cal U}_{VRH}={\cal U}_{hl}$
yields the vertical line separating the VRH and half-loop regions.

Transport in flux arrays with columnar pins is described by
a phase diagrams similar to that of Fig. 1.
The difference for planar disorder is that
transport is one dimensional in this case
and rare fluctuations 
in the spatial distribution
of twins can impede VRH and dominate
flux-line dynamics.
The vortex line can encounter a 
region where
no favorable twins are available at the distance of the optimal jump.
The vortex will then remain trapped in this region for a long time
and the resistivity can be greatly suppressed.
Rare fluctuations can also occur in samples with
columnar pins, but in that case 
they will dominate transport
only at extremely small fields, when the number of rare regions exceeds
the number of vortices.  

At a given temperature and for applied currents below $J_L$, 
a vortex can jump from one twin plane 
to another 
at a distance $y$ if the energy difference per unit length between the 
initial and final configuration is within a range 
$\Delta \epsilon\sim E_ky/Ld$.
A trap is then a region of configuration space $(y,\epsilon$)
void of localized states within a spatial distance $y$ and  an energy band $\Delta\epsilon$ around
the initial vortex state. A vortex that has entered such a trap or ``break''
will remain in the trap for a time 
$t_w\approx t_0\exp(2y/l_{\perp})$,
where $l_{\perp}=T/\sqrt{2\epst U}$ is the localization length and
$t_0$ a microscopic time scale.
The probability
of finding such a break is given by a Poisson distribution,
$P(y)\approx P_0(y)\exp[-Ag(\mu)y\Delta\epsilon]$,
where $P_0(y)$ is the concentration of localized states in the energy band
$\Delta\epsilon$, $P_0(y)\approx 2Ag(\mu)\Delta\epsilon$ and $A\sim 1$ 
a numerical constant. The mean waiting time between jumps
is 
\begin{equation}
\label{eq:meant}
\overline{t_w}\approx\int_0^{\infty} dy P(y) t_0e^{2y/l_{\perp}(T)}.
\end{equation}
For $L>>(T/E_k)^2\alpha w_k=L^*$,
the integral can be evaluated at the saddle point, corresponding to the
situation where the mean waiting time is controlled by ``optimal breaks'' 
of transverse witdth $y_{l}^*\approx l_{\perp} L/L^*$,
with the result,
$\overline{t_w}\sim t_0 \sqrt{L/L^*}e^{L/L^*}$.
The optimal breaks correspond to the longest trapping time
and are most effective at preventing flux motion.
The inverse of the trapping time determines the characteristic 
rate of jumps,
yielding a linear resistivity,

\begin{equation}
\label{eq: resbreakl}
\rho_{bl}\approx\rho_0(T/{\cal U}_{Mott})e^{-({\cal U}_{Mott}/T)^2}.
\end{equation}

For currents above $J_L$ the typical energy per unit length available
to a flux line for jumping a distance $y$ is $\Delta\epsilon\sim f_Ly$.
Again, for $J<<(J_1/\alpha)(E_k/T)^2=J^*$
the main contribution to the resistivity is from
``optimal traps'' of transverse size
$y^*_b\approx l_{\perp}(J^*/J)$,
with the result,
\begin{equation}
\label{eq: resbreak}
\rho_b\approx\rho_0(T/{\cal U}_{VRH})e^{-({\cal U}_{VRH}/T)^2}. 
\end{equation}
The contribution
to the resistivity from tunneling \`a la Mott (both in the linear and nonlinear
regimes) always dominates that from hopping between
rare optimal traps
if both mechanisms of transport can occur.
On the other hand,
in one dimension if the sample is wide enough in the direction of
flux-line motion to contain optimal traps,
tunneling \`a la Mott
simply cannot take place because flux lines cannot get around the traps.
These rare traps with large waiting times will then control
transport. 
If $W$ is the sample width in the $y$ direction, the condition
for having optimal traps of width $y^*_{l,b}$ is $P(y^*_{l,b})W>1$.
\begin{figure}
\epsfxsize=3.0truein
\hskip 0.5truein \epsffile{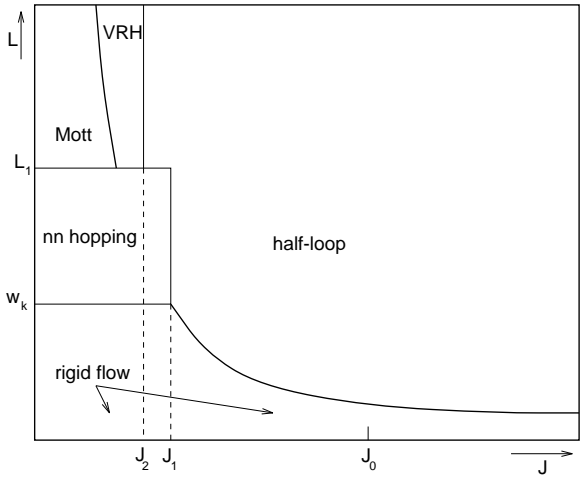}
\caption{The $(L,J)$ phase diagram for $\alpha=g(\mu)dU<E_k/T$.}
\label{phasetwo}
\end{figure}

\noindent Optimal traps will be present only if $J>J_w=J^*/\ln(2W/l_{\perp})$
for $J>J_L$ and if $L<L_w=L^*\ln(2W/l_{\perp})$ for $J<J_L$.

A flux line can, however, escape a break by
nucleating a half-loop, i.e., by tunneling directly into 
conduction band, if $y_{hl}>y^*_b$, or $\alpha<E_k/T$.
The only effect of rare fluctuations in this case is of suppressing
VRH for $J>J_w$,
extending to lower currents the region where transport occurs
via half-loop nucleation. For instance 
the right (high current) boundary
of the VRH region in
the phase diagram of Fig. 1 will be pushed down to $J_w$
if $J_w<J_2$, or $\ln(2W/y_T)>(E_k/\alpha T)^2$.
Similar considerations apply to the linear response.
Only if $\alpha>E_k/T$, there will be a portion of the $(J,L)$
plane where breaks 
dominate transport, as shown in Fig. 2.
For YBCO, we estimate $E_k\sim 1K\AA^{-1} d$. 
Assuming $\alpha\sim U/\gamma$, the condition $\alpha>E_k/T$
can only be satisfied at low fields ($B<1KG$ for $d\sim 200\AA$).
The sample will contain optimal breaks if $W>30\AA\exp(J^*/J)$,
with $J^*\sim 4\times 10^5$Amp/cm$^2$ at $80K$.

If the sample is too short to contain optimal breaks,
i.e., $W P(y\sim y^*_{l,b})<1$, the dynamics
is controlled  by the trap with the longest waiting time,
$t(y_f)\sim \exp(y_f/l_{\perp})$, with $y_f$ 
determined by the condition $W P(y_f)\sim 1$. The corresponding 
resistivity is 
$\rho_W\approx\rho_0e^{-y_f/l_{\perp}}$.
In this case the relevant physical quantity is the logarithm 
of the resistivity,
\begin{eqnarray}
\label{eq:reslog}
\ln(& &\rho_W/\rho_0)=-y_f/l_{\perp}\nonumber\\
& &\approx - {{\cal U}_{VRH}\over T}
  \Big\{\ln\Big[{2W\over l_{\perp}}{T\over {\cal U}_{VRH}}
     \Big(\ln(2W/ l_{\perp})
  \Big)^{1/2}\Big]\Big\}^{1/2}. 
\end{eqnarray}
The leading dependence of Eq. \ref{eq:reslog} on current and temperature
is the same as that of the VRH contribution.
Equation (\ref{eq:reslog}) also contains, however, logarithmic terms that
in sufficiently short samples will give a random spread of values
of the resistivity from sample to sample. These effects 
have been discussed for semiconductors \cite{ruzin}. 
In this case a more relevant
physical quantity rather than the resistivity itself is the 
distribution of the logarithms of the resistivity over different samples.
The expression (\ref{eq:reslog}) determines the position of the maximum of
this distribution.

\vskip .2in
This work was supported by the National Science Foundation through 
Grant No. DMR91-12330 and through the U.S. Department of Energy, 
BES-Material Sciences, under contract No. W-31-109-ENG-38.
MCM is grateful to Argonne National Laboratory and to the Institute for 
Scientific
Interchange in Torino, Italy, for hospitality and partial support
during the completion of this work.
VMV thanks Gianni Blatter and the Swiss
National Foundation for supporting his visit at ETH-Z\"urich were
part of this work was carried out.
Finally, we both thank David Nelson for many stimulating discussions.

% tables


\begin{thebibliography}{99}
\bibitem{notenn}
When $\alpha<1$ nn hopping can occur only if
the chemical potential $\mu$ falls in the tails of the impurity band,
so that $g(\mu)d<1/\gamma$. Only then $\gamma<U$ and $L_1>w_k$.
\end{thebibliography}
\end{document}